\documentclass[11pt]{article}
\usepackage[textwidth=15.2cm,textheight=22cm]{geometry}
\usepackage{amsmath,amssymb}
\usepackage{latexsym}
\usepackage{graphicx}
\usepackage{bbm}
\usepackage{bm}
\usepackage{cite}
\allowdisplaybreaks[1]
\numberwithin{equation}{section}

\newcommand{\tr}{{\rm tr}}

\newcommand{\be}{\begin{equation}}
\newcommand{\ee}{\end{equation}}
\newcommand{\ba}{\begin{eqnarray}}
\newcommand{\ea}{\end{eqnarray}}
\newcommand{\bdm}{\begin{displaymath}}
\newcommand{\edm}{\end{displaymath}}

\newcommand\fr[1]{\frac{1}{#1}}

\def\p{\partial}
\def\bp{\bar \partial}
\def\ba{\bar A}

\def\bc{\bar C}
\def\bh{\bar h}
\def\beq{\begin{equation}}
\def\eeq{\end{equation}}

\newcommand{\calN}{{\mathcal N}}

\DeclareMathAlphabet{\mathpzc}{OT1}{pzc}{m}{it}

\newcommand{\ndt}{\noindent}

\def\bea{\begin{eqnarray}}
\def\eea{\end{eqnarray}}
\def\beas{\begin{eqnarray*}}
\def\eeas{\end{eqnarray*}}
\def\sla{\raise.15ex\hbox{$/$}\kern-.57em}

\def\parm{{\partial}_{-}}
\def\parp{\partial_+}

\def\lt{\tilde{\lambda}}
\def\spa#1.#2{\left\langle#1\,#2\right\rangle}
\def\spb#1.#2{\left[#1\,#2\right]}

\begin{document}

\begin{titlepage}
\begin{flushright}    
{\small $\,$}
\end{flushright}
\vskip 1cm
\centerline{\Large{\bf {KLT relations from the Einstein-Hilbert Lagrangian}}}
\vskip 1.5cm
\centerline{Sudarshan Ananth and Stefan Theisen}
\vskip .5cm
\centerline{\it {Max-Planck-Institut f\"{u}r 
Gravitationsphysik}}
\centerline{\it {Albert-Einstein-Institut, Golm, Germany}}
\vskip 1.5cm
\centerline{\bf {Abstract}}
\vskip .5cm
\ndt The Kawai-Lewellen-Tye (KLT) relations derived from string theory tell us that perturbative gravity amplitudes are the ``square" of the corresponding amplitudes in gauge theory. Starting from the light-cone Lagrangian for pure gravity we make these relations manifest off-shell, for three- and four-graviton vertices, at the level of the action.
\vfill
\end{titlepage}

\section{Introduction}

The Kawai-Lewellen-Tye (KLT) relations relate tree-level amplitudes in closed and open string theories~\cite{KLT}. In the field theory limit the KLT relations, for three- and four-point amplitudes, reduce to\footnote{For higher-point generalizations see~\cite{BDPR}.}
\bea
\label{kltrelate}
M^{\rm tree}_3(1,2,3)&&\!\!\!\!\!\!\!\!\!\!=A^{\rm tree}_3(1,2,3)\,A^{\rm tree}_3(1,2,3)\ , \nonumber \\
M^{\rm tree}_4(1,2,3,4)&&\!\!\!\!\!\!\!\!\!\!=-i\,s_{12}\,A^{\rm tree}_4(1,2,3,4)\,A^{\rm tree}_4(1,2,4,3)\ ,
\eea
where the $M_n$ represent gravity amplitudes and the $A_n$ are color-ordered~\cite{shf,dixon} amplitudes in pure Yang-Mills theory ($s_{ij}\equiv-(p_i+p_j)^2$). Although the KLT relations apply only at the tree-level they have been used, with great success, in conjunction with unitarity based methods to derive loop amplitudes in gravity~\cite{BDPR,BDPR2}. In particular, these relations have proven invaluable in studying the ultra-violet properties of $\calN=8$ supergravity~\cite{finite}. The question of whether the KLT relations are valid only for on-shell amplitudes or, more generally, at the level of the Lagrangian remains open~\cite{living}. This is the issue we focus on in this letter.

The tree-level amplitudes take a very compact form in a helicity basis. Thus when attempting to derive the KLT relations starting from the gravity Lagrangian it seems natural to work in light-cone gauge where only the helicity states propagate. Tree-level amplitudes in which precisely two external legs carry negative helicity are called maximally helicity violating (MHV) amplitudes. A very simple expression for all the MHV amplitudes in Yang-Mills theory was given in~\cite{PT}. An MHV-Lagrangian (also referred to as the CSW Lagrangian) where the fundamental vertices are off-shell versions of the MHV amplitudes was proposed in~\cite{CSW}. In~\cite{Rosly} and~\cite{Mansfield} it was shown how this MHV-Lagrangian can be derived from the usual light-cone Yang-Mills Lagrangian by a suitable field redefinition.

In this letter we perform a field redefinition, similar to that in~\cite{Rosly,Mansfield}, on the light-cone gravity Lagrangian. Although the shifted Lagrangian is not simply the sum of MHV-vertices, the off-shell KLT relations, to the order examined in this letter, are manifest. 

\section{Yang-Mills}

We start by sketching schematically, the proposal of~\cite{Rosly,Mansfield} for Yang-Mills. The light-cone Yang-Mills Lagrangian is of the form
\bea
\label{schemym}
{\it L}\,\sim\,{\it L}_{+-}+{\it L}_{++-}+{\it L}_{+--}+{\it L}_{++--}\ ,
\eea
where the indices, in no particular order, refer to helicity. The field redefinition maps the first two terms (the kinetic and one cubic term) into a purely kinetic term. This transformation also generates an infinite series of higher order terms producing exactly the MHV-Lagrangian
\bea
{\it L}_{\,YM}\,\sim\,{\it L}_{+-}+{\it L}_{+--}+{\it L}_{++--}+{\it L}_{+++--}+{\it L}_{++++--}+\ldots+{\it L}_{(+)^n--}+\ldots\ .
\eea
Again, this is merely a formal way of writing the Lagrangian. For example, ${\it L}_{++--}$ receives contributions from the two inequivalent orderings $\tr(A\ba A\ba)$ and $\tr(AA\ba\ba)$ where $A$ and $\ba$ are gluons of helicity\footnote{The helicity label assumes that the particle is outgoing.} $+1$ and $-1$ respectively. Each trace is multiplied by an off-shell continuation (cf. appendix A) of the appropriate Parke-Taylor amplitude~\cite{PT,dixon}
\bea
\frac{\spa{k}.{l}^4}{\prod_{i=1}^n\,\spa{i}.{(i+1)}}\ ,\qquad \mbox {\footnotesize $n+1\equiv 1$}\ .
\eea
We will not go into details regarding the derivation of these results which can be found in~\cite{Rosly,Mansfield,EM,QMC}. The analysis in the gravity case is completely analogous and is presented in detail in section 3. The hope is that a similar field redefinition in pure gravity will generate interaction terms which make KLT factorization manifest. The purpose of this letter is to examine this issue.

\section{Gravity in light-cone gauge}

We follow closely, in this section, the light-cone formulation of gravity in~\cite{ABHS}. Here, we only review the key features of this formulation and refer the reader to appendix C in~\cite{ABHS} for a detailed derivation of the results presented below.$\\$
\ndt The Einstein-Hilbert action reads
\bea
S_{EH}=\int\,{d^4}x\,{\cal L}\,=\,\frac{1}{2\,\kappa^2}\,\int\,{d^4}x\,{\sqrt {-g}}\,R\ ,
\eea
where $g=\det{g_{\mu\nu}}$ and $R$ is the curvature scalar. Light-cone gauge is chosen by setting
\bea
g_{--}\,=\,g_{-i}\,=\,0\ ,\qquad \mbox {\footnotesize $i=1,2$}\ .
\eea
Our conventions and notation are explained in appendix A. The metric is parameterized as follows
\bea
g_{+-}\,=\,-\,e^\frac{\psi}{2}\ ,\quad g_{ij}\,=\,e^\psi\,\gamma_{ij}\ .
\eea
The field $\psi$ is real while $\gamma_{ij}$ is a $2\times 2$ real, symmetric, unimodular matrix. The $R_{-i}=0$ constraint allows us to eliminate $g^{-i}$. From the $R_{--}=0$ constraint we find 
\bea
\psi=\fr{4}\,\fr{\parm^2}\,\big(\parm \gamma^{i j} \parm \gamma_{ij}\big)\ .
\eea
The Lagrangian density now reads
\bea
{\cal L}=\fr{2\kappa^2}\,\sqrt{-g}\;\bigg( 2 g^{+-} R_{+-} +g^{ij} R_{ij}\bigg)\ .
\eea
We expand this to find~\cite{lcgref}
\bea
\label{finallag}
{\cal L}&=&\fr{2\kappa^2}\bigg\{\;\mathrm{e}^{\psi}\bigg(\frac{3}{2}\parp\parm\psi-\fr{2}\parp\gamma^{ij}\parm\gamma_{ij}\bigg) \nonumber \\
&&-\mathrm{e}^{\frac{\psi}{2}}\gamma^{ij}\bigg(\fr{2}\p_i\p_j\psi-\frac{3}{8}\p_i\psi\p_j\psi-\fr{4}\p_i\gamma^{kl}\p_j\gamma_{kl}+\fr{2}\p_i\gamma^{kl}\p_k\gamma_{jl}\bigg) \nonumber \\
&&-\fr{2}\mathrm{e}^{-\frac{3}{2}\psi}\gamma^{ij}\fr{\parm}R_i\fr{\parm}R_j\,\Bigg\}\ ,
\eea
where
\bea
R_i=\mathrm{e}^{\psi}\bigg(-\fr{2}\parm\gamma^{jk}\p_i\gamma_{jk}+\frac{3}{2}\parm\p_i\psi-\fr{2}\p_i\psi\parm\psi\bigg)-\p_k\bigg(\mathrm{e}^{\psi}\gamma^{jk}\parm\gamma_{ij}\bigg)\ .
\eea
This is the closed form of the Lagrangian.

\subsection{The perturbative expansion}

In order to obtain a perturbative expansion of the metric we choose 
\bea
\gamma_{ij}=\left(\mathrm{e}^{\kappa H}\right)_{ij}\ ,\qquad H=\fr{\sqrt 2}\begin{pmatrix} h+\bh & -i(h-\bh)\\-i(h-\bh) &-h-\bh\end{pmatrix}\ ,
\eea
where $h$ and $\bh$ represent gravitons of helicity $+2$ and $-2$ respectively. The light-cone Lagrangian density for pure gravity, to order $\kappa^2$~\cite{ABHS}, reads~\footnote{\ndt As seen in appendix C of~\cite{ABHS}, a field redefinition which removes occurences of $\parp$ from the interaction terms has been performed.}
\bea
\label{lcg}
{\cal L}\;\;=&&\!\!\!\!\!\!\bh\, \square \, h \nonumber \\
&&\!\!\!\!\!\!\!\!+2\kappa\,\bh\,\parm^2\bigg(\frac{\bp}{\parm}h\frac{\bp}{\parm}h-h\frac{\bp^2}{\parm^2}h\bigg)+2\kappa\,h\,\parm^2\bigg(\frac{\p}{\parm}\bh\frac{\p}{\parm}\bh-\bh\frac{\p^2}{\parm^2}\bh\bigg) \nonumber \\
&&\!\!\!\!\!\!\!\!\, \nonumber \\
&&\!\!\!\!\!\!\!\!+2\kappa^2{\biggl \{}\,\fr{\parm^2}\big(\parm h\parm\bh\big)\frac{\p\bp}{\parm^2}\big(\parm h\parm\bh\big)+\fr{\parm^3}\big(\parm h\parm\bh\big)\left(\p\bp h\,\parm\bh+\parm h\p\bp\bh\right) \nonumber\\
&&\!\!\!\!\!\!\!\!-\fr{\parm^2}\big(\parm h\parm\bh\big)\,\left(2\,\p\bp h\,\bh+2\,h\p\bp\bh+9\,\bp h\p\bh+\p h\bp\bh-\frac{\p\bp}{\parm}h\,\parm\bh-\parm h\frac{\p\bp}{\parm}\bh\right) \nonumber  \\
&&\!\!\!\!\!\!\!\!-2\fr{\parm}\big(2\bp h\,\parm\bh+h\parm\bp\bh-\parm\bp h\bh\big)\,h\,\p\bh-2\fr{\parm}\big(2\parm h\,\p\bh+\parm\p h\,\bh-h\parm\p\bh\big)\,\bp h\,\bh \nonumber \\
&&\!\!\!\!\!\!\!\!-\fr{\parm}\big(2\bp h\,\parm\bh+h\parm\bp\bh-\parm\bp h\bh\big)\fr{\parm}\big(2\parm h\,\p\bh+\parm\p h\,\bh-h\parm\p\bh\big) \nonumber \\
&&\!\!\!\!\!\!\!\!-h\,\bh\,\bigg(\p\bp h\,\bh+h\p\bp\bh+2\,\bp h\p\bh+3\frac{\p\bp}{\parm}h\,\parm\bh+3\parm h\frac{\p\bp}{\parm}\bh\bigg){\biggr \}}\ .
\eea
As in (\ref {schemym}) the three-vertex terms are of the form $(-,+,+)$ and $(+,-,-)$. In analogy to Yang-Mills, a solution to the self-duality condition
\bea
R_{\mu\nu\rho\sigma}=\frac{i}{2}\,\epsilon_{\mu\nu}{}^{\alpha\beta}R_{\alpha\beta\rho\sigma}\ ,
\eea
is 
\bea
{\bar h}=0\ ,\quad \square\,h+2\kappa\,\parm^2 \bigg(\frac{\bp}{\parm} h \frac{\bp}{\parm} h-h \frac{\bp^2}{\parm^2} h \bigg)\,=\,0\ ,
\eea
where the second relation is the $\bh$ equation of motion (at $\bh=0$). Thus, as in Yang-Mills, we will map the first two terms in (\ref {lcg}) to a free theory. Further discussions regarding this point may be found in~\cite{EM}.

\subsection{The field redefinition}

We seek a transformation $(h,\bh)\rightarrow(C,\bc)$ such that\footnote{Note that the d'Alembertian is $\square=2(\p\bp-\parp\parm)$. See appendix A for further details.}
\bea
{\it K}=-\bh\parp\parm h+\bh V(h)=-\bc\parp\parm C+\bc\p\bp C\ ,
\eea
where
\bea
V(h)=\p\bp h+\kappa\,{\parm^2}\bigg(\frac{\bp}{\parm}h\,\frac{\bp}{\parm}h-h\,\frac{\bp^2}{\parm^2}h\bigg)\ .
\eea
The remaining three- and four-point vertices in (\ref {lcg}) all involve exactly two negative helicity gravitons. Since MHV amplitudes also involve exactly two negative helicity legs, we aim to preserve this structure\footnote{We point out that higher order terms in (\ref {finallag}) do not possess this structure.}. In analogy with Yang-Mills, we choose $h$ to be a function of $C$ alone while $\bh$ is chosen to be a function of both $C$ and $\bc$. This field redefinition is not unique and we will comment on this below. 

To find the explicit transformation, which is in fact a canonical transformation on the phase space with coordinates $(C,\pi_C)$, we start with a generating function of the form $G(C,\pi_h)=\int g(C)\,\pi_h$. Then
\bea
\pi_C\equiv\frac{\partial {\it L}}{\partial(\parp C)}=\parm\bc=\frac{\delta G}{\delta C}=\int\frac{\delta g}{\delta C}\pi_h\ ,\qquad h=\frac{\delta G}{\delta \pi_h}=g(C)\ .
\eea
Since $\pi_h=\parm\bh$ we have
\bea
\label{relate2}
\parm\bc(y)=\int d^3x\,\parm\bh(x)\;\frac{\delta h(x)}{\delta C(y)}\ ,
\eea
where the integral is performed on a surface of constant $x^+$. The Lagrangian density then reads (here and below we drop surface terms)
\bea
{\cal L}=-\bc\parp\parm C+\bc\p\bp C&&\!\!\!\!\!\!\!\!=\parm\bc\,\parp C-\parm\bc\,\frac{\p\bp}{\parm}C\ .
\eea
Using (\ref {relate2}) the Lagrangian becomes
\bea
L=\int d^3x\,\parm\bh(x)\parp h(x)-\int d^3x\,\int d^3y\,\parm\bh(y)\frac{\p\bp}{\parm}C(x)\,\frac{\delta h(y)}{\delta C(x)}\ .
\eea
We want this to be equal to
\bea
L=\int d^3x\,\bigg(\parm\bh(x)\parp h(x)-\parm\bh(x)\fr{\parm}V(h(x))\bigg)\ ,
\eea
implying that
\bea
\frac{\p\bp}{\parm}h(x)+\kappa\,\parm\,\left(\frac{\bp}{\parm}h\,\frac{\bp}{\parm}h\,-\,h\,\frac{\bp^2}{\parm^2}h\right)(x)
=\int d^3y\,\frac{\p\bp}{\parm}C(y)\frac{\delta h(x)}{\delta C(y)}\ .
\eea
In momentum space, this becomes
\bea
\label{checker2}
&&\frac{p{\bar p}}{p_-}h(p_-)\,-\,\int d^3m\,\frac{m{\bar m}}{m_-}C(m)\,\frac{\delta h(p)}{\delta C(m)}\,= \nonumber \\
&&-\,\kappa\,\int d^3k\,d^3l\;\delta^{(3)}(p-k-l)\;(k_-+l_-)\bigg(\frac{\bar k\bar l}{k_-l_-}-\frac{{\bar l}^2}{l_-^2}\bigg)\,h(k)\,h(l)\ .
\eea
For $h$, we choose the ansatz
\bea
\label{ansh}
h(p)=\sum_{n=1}^\infty\,\int\,\prod_{i=1}^n\,d^3k_i\;Z^{(n)}(p_1,k_1,\ldots,k_n)\,C(k_1)\ldots C(k_n)\ ,
\eea
so (\ref {checker2}) implies
\bea
&&\int d^3k\,d^3l\,\bigg(\frac{p{\bar p}}{p_-}\,-\,\frac{k{\bar k}}{k_-}\,-\,\frac{l{\bar l}}{l_-}\bigg)\;Z^{(2)}(p,k,l)\,C(k)\,C(l)\,= \nonumber \\
&&-\kappa\,\int d^3k\,d^3l\,(k_-+l_-)\,\bigg(\,\frac{{\bar k}{\bar l}}{k_-l_-}\,-\,\frac{{\bar l}^2}{l_-^2}\,\bigg)\,C(k)\,C(l)\,\delta^{(3)}(p-k-l)\ .
\eea
Thus
\bea
Z^{(1)}(p,k)&&\!\!\!\!\!\!=\,\delta^{(3)}(p-k)\ , \nonumber \\
Z^{(2)}(p,k,l)&&\!\!\!\!\!\!=\,\frac{\kappa}{2}\;(k_-+l_-)\,\frac{\frac{{\bar l}^2}{l_-^2}\,+\,\frac{{\bar k}^2}{k_-^2}\,-\,2\frac{{\bar k}{\bar l}}{k_-l_-}}{\frac{p{\bar p}}{p_-}\,-\,\frac{k{\bar k}}{k_-}\,-\,\frac{l{\bar l}}{l_-}}\,\delta^{(3)}(p-k-l) \nonumber \\
&&\!\!\!\!\!\!=\,-\frac{\kappa}{2}\;\frac{p_-^2}{k_-l_-}\,\frac{\spb{k}.{l}}{\spa{k}.{l}}\,\delta^{(3)}(p-k-l)\ .
\eea
From (\ref {relate2}) we also find
\bea
p_-\bh(p)=p_-{\bar C}(p)-\int d^3k\,d^3l\,k_-\big(\,Z^{(2)}(-k,-p,l)+Z^{(2)}(-k,l,-p)\big){\bar C}(k)C(l)+\ldots
\eea
which can be rewritten as
\bea
\label{ansbh}
\bh(p)\,=\,{\bar C}(p)+\kappa\,\int d^3k\,d^3l\,\frac{k_-^3}{p_-^2l_-}\,\frac{\spb{k}.{l}}{\spa{k}.{l}}\,{\bar C}(k)C(l)\,+\,\ldots
\eea
It is straightforward to work out a recursion relation for the coefficients $Z^{(n)}$ which can then be solved to any desired order. We will not present the details here.

\vskip 0.5cm
\subsection{The shifted gravity action}
\vskip 0.3cm

After performing the field redefinition described in the previous section we find that the gravity action, to order $\kappa^2$, is
\bea
\label{shifgravold}
&&\!\!\!\!\!\!\!\!\int d^4p\;\bc(-p)\,p^2\,C(p)+\kappa\int d^4p\;d^4k\;d^4l\;\frac{\spa{k}.{l}^6}{\spa{l}.{p}^2\spa{p}.{k}^2}\,C(p)\bc(k)\bc(l)\;\delta^{(4)}(p\!+\!k\!+\!l) \\
+&&\!\!\!\!\!\!\!\!\kappa^2\int d^4p\;d^4q\;d^4k\;d^4l\;\frac{\spa{k}.{l}^8\spb{k}.{l}}{\spa{k}.{l}\spa{k}.{p}\spa{k}.{q}\spa{l}.{p}\spa{l}.{q}\spa{p}.{q}^2}\,C(p)C(q)\bc(k)\bc(l)\;\delta^{(4)}(p\!+\!q\!+\!k\!+\!l) \nonumber \\
\;&& \;\nonumber \\
+&&\!\!\!\!\!\!\!\!\kappa^2\int d^4p\,d^4q\,d^4k\,d^4l\left(J(p,q,k,l)\,p^2+K(p,q,k,l)\,k^2\right)C(p)C(q)\bc(k)\bc(l)\delta^{(4)}(p\!+\!q\!+\!k\!+\!l)\ .\nonumber
\eea We stress that the coefficients in the action above are off-shell. Note that the four-graviton amplitude does not receive exchange contributions due to the structure of the action at the cubic level after the field redefinitions (\ref {ansh}) and (\ref {ansbh}). The functions $J$ and $K$ turn out to be fairly complicated but are irrelevant for on-shell four-point scattering since the third line vanishes on-shell. In particular, when interaction vertices are proportional to the free equations of motion they can be eliminated by a suitable field redefinition~\cite{redef}. The required field redefinitions are\footnote{This field redefinition changes the structure $h=h(C)$ to $h=h(C,\bc)$ but affects only four- and higher-point vertices. This demonstrates the non-uniqueness of the field redefinition in section 3.2.}
\bea
\label{further}
\!\!\!\!\!\!\!\!\!\!\!\!\!\!\!\!\!\!C(p)\rightarrow&&\!\!\!\!\!\!C(p)-\!\kappa^2\int\!d^4q\,d^4k\,d^4l\;\;K(k,q,-p,l)\;C(k)C(q)\bc(l)\;\delta^{(4)}(-p\!+\!q\!+\!k\!+\!l) \ , \nonumber \\
\!\!\!\!\!\!\!\!\!\!\!\!\!\!\!\!\!\!\bc(p)\rightarrow&&\!\!\!\!\!\!\bc(p)-\!\kappa^2\int\!d^4q\,d^4k\,d^4l\;\;J(-p,q,k,l)\;\bc(k)\bc(l)\,C(q)\;\delta^{(4)}(-p\!+\!q\!+\!k\!+\!l) \ ,
\eea
and these eliminate the third line in (\ref {shifgravold}). The light-cone action for gravity to order $\kappa^2$ thus reads
\bea
\label{shifgrav}
&&\!\!\!\!\!\!\!\!\int d^4p\;\bc(-p)\,p^2\,C(p)+\kappa\int d^4p\;d^4k\;d^4l\;\frac{\spa{k}.{l}^6}{\spa{l}.{p}^2\spa{p}.{k}^2}\,C(p)\bc(k)\bc(l)\;\delta^{(4)}(p\!+\!k\!+\!l) \\
+&&\!\!\!\!\!\!\!\!\kappa^2\int\! d^4p\;d^4q\;d^4k\;d^4l\;\frac{\spa{k}.{l}^8\spb{k}.{l}}{\spa{k}.{l}\spa{k}.{p}\spa{k}.{q}\spa{l}.{p}\spa{l}.{q}\spa{p}.{q}^2}\,C(p)C(q)\bc(k)\bc(l)\,\delta^{(4)}(p\!+\!q\!+\!k\!+\!l)\ . \nonumber
\eea
These off-shell vertices clearly factorize into products of off-shell MHV vertices in Yang-Mills. In particular this confirms, {\it {off-shell}}, the relations (\ref {kltrelate}) for three- and four-point vertices. It will be interesting to see if this KLT factorization extends to higher orders in the action where non-MHV vertices appear.

\vskip 0.5 cm
\begin{center}
* ~ * ~ *
\end{center}
\vskip -0.1 cm

In contrast to the Yang-Mills case, the MHV vertices in gravity appear only after a further field redefinition (\ref {further}) that removes interaction vertices proportional to the free equations of motion. This was to be expected given that the gravity Lagrangian, unlike Yang-Mills, does not stop at quartic order and that the MHV gravity amplitudes are non-holomorphic~\cite{BGK}. Furthermore MHV vertices in the gravity Lagrangian are not sufficient to compute all the non-MHV diagrams, at least for our choice of field variables. For example the 5-point amplitude $M^{\rm tree}(+,+,-,-,-)$ has contributions from the MHV vertices but also from a direct contact vertex present in the original Lagrangian\footnote{In~\cite{bohr}, the five-point non-MHV graph is simply a sum of MHV-exchange diagrams. In our case there is also a direct contribution: this is not surprising since, in our Lagrangian, we have eliminated the three-vertex $M(+,+,-)$ and so do not have a contribution equivalent to $D_2$ in equation (3.14) of that reference.}. The five-point MHV amplitude $M^{\rm tree}(+,+,+,-,-)$ is special in that it has three contributions: one term from the original Lagrangian and two from the field redefinition acting on the three- and four-point vertices. Otherwise, as in Yang-Mills, all $n$-point ($n>5$) MHV amplitudes are generated by the field redefinitions alone.

The discussion in the main body of this letter dealt with light-cone gravity at tree-level. At the loop level, field redefinitions have to be considered with much greater care. If the Jacobian of the field redefinition is not unity it will lead to additional interaction terms~\cite{redef}. Even if the Jacobian is classically one there may be anomalies which lead to additional interaction terms as proposed in the context of the MHV Lagrangian for Yang-Mills in~\cite{Rosly}; see also the discussion in~\cite{EM,QMC}. 

An interesting question is whether the Lagrangians of $\calN=8$ supergravity and $\calN=4$ superYang-Mills share a similar relationship. Since there exist superfield formulations, in light-cone gauge, for both these theories~\cite{sff} a similar analysis is certainly worth performing.

\subsection*{Acknowledgments}

We thank Lars Brink, Hermann Nicolai, Alexei Rosly, Adam Schwimmer and Hidehiko Shimada for discussions. 

\appendix 
\section{Conventions and notation}

We work with the metric $(-,+,+,+)$ and define
\be
x^\pm\,=\,\fr{\sqrt 2}\,(x^0\,\pm\,x^3)\ , \quad \partial_\pm\,=\,\fr{\sqrt 2}\,(\partial_0\,\pm\,\partial_3)\ .
\ee
$x^+$ plays the role of light-cone time and $\partial_+$ the light-cone Hamiltonian. $\parm$ is now a spatial derivative and its inverse, $\frac{1}{\parm}$, is defined using the prescription in~\cite{SM}. We define
\bea
x&&\!\!\!\!\!\!\!\!=\fr{\sqrt 2}\,(x^1\,+i\,x^2)\ , \quad {\bar \partial}\equiv\frac{\partial}{\partial x}=\fr{\sqrt 2}\,(\partial_1\,-\,i\,\partial_2)\ , \nonumber \\
{\bar x}&&\!\!\!\!\!\!\!\!=\fr{\sqrt 2}\,(x^1\,-i\,x^2)\ , \quad \partial\equiv\frac{\partial}{\partial {\bar x}}=\fr{\sqrt 2}\,(\partial_1\,+\,i\,\partial_2)\ .
\eea
\vskip -0.1cm
\ndt A four-vector $p_\mu$ may be expressed as a bispinor $p_{a \dot{a}}$ using the $\sigma^\mu=(-{\bf {1}},\,{\bf {\sigma}})$ matrices 
\beq
\label{paad}
p_{a \dot{a}}\,\equiv\,p_\mu\,{(\sigma^\mu)}_{a \dot{a}}\,=\,\left( 
 \begin{matrix}
-p_0+p_3 & \;p_1-ip_2\, \\ p_1+ip_2 & \;-p_0-p_3\, 
\end{matrix}
\right)=\sqrt{2}
\left( 
 \begin{matrix}
-p_- & \;{\overline p}\, \\ p & \;-p_+\, 
\end{matrix}
\right)
\ . 
\eeq
The determinant of this matrix is
\bea
{\mbox {det}}\,(\,p_{a \dot{a}}\,)\,=\,-2\,(\,p{\overline p}-p_+p_-\,)\,=\,-\,p^\mu p_\mu\ .
\eea
When the vector $p_\mu$ is light-like we have $p_+\,=\,\frac{p{\overline p}}{p_-}$ which is the on-shell condition. We then define holomorphic and anti-holomorphic spinors~\footnote{When working with a Lorentzian signature, choosing $\lt_{\dot{a}}=\pm(\lambda_a)^*$ ensures that $p_{a \dot{a}}$ is real.}
\beq
\lambda_{a}\,=\,\frac{2^\fr{4}}{\sqrt p_-} 
\left( 
\begin{matrix} 
p_-  \\ 
\, -p 
\end{matrix}
\right) 
\ , 
\qquad 
\lt_{\dot{a}}\,=\,-(\lambda_a)^*\,=\,-\,\frac{2^\fr{4}}{\sqrt p_-} 
\left( 
\begin{matrix} 
p_- \\ 
\, -{\overline p} 
\end{matrix}
\right) 
\ , 
\eeq
such that $\lambda_{a}\lt_{\dot{a}}$ agrees with (\ref {paad}) on-shell. We define the off-shell holomorphic and anti-holomorphic spinor products~\cite{QMC}
\be \label{product}
\spa{i}.{j}\,=\,\sqrt{2}\,\frac{p^i\,p_-^j\,-\,p^j\,p_-^i}{\sqrt{p_-^i\,p_-^j}}\ ,\qquad \spb{i}.{j}\,=\,\sqrt{2}\,\frac{{\bar p}^i\,p_-^j\,-\,{\bar p}^j\,p_-^i}{\sqrt{p_-^i\,p_-^j}}\ .
\ee
\vskip -0.3cm
\ndt Their product is
\bea
\spa{i}.{j}\spb{j}.{i}=s_{ij}\equiv-(p_i+p_j)^2\ .
\eea


\begin{thebibliography}{Ref}
\bibitem{KLT}{H. Kawai, D.C. Lewellen and S.H.H. Tye, {\it Nucl. Phys.} {\bf B 269} (1986) 001.}
\bibitem{BDPR}{Z. Bern, L. J. Dixon, M. Perelstein and J.S. Rozowsky, {\it Nucl. Phys.} {\bf B 546} (1999) 423, {\tt hep-th}/9811140.}
\bibitem{shf}{F. A. Berends and W. Giele, {\it Nucl. Phys.} {\bf B 294} (1987) 700.}
\bibitem{dixon}{L. Dixon, TASI (1995) 539, {\tt hep-ph}/9601359.}
\bibitem{BDPR2}{Z. Bern, L. J. Dixon, M. Perelstein and J. S. Rozowsky, {\it Phys. Lett.} {\bf B 444} (1998) 273, {\tt hep-th}/9809160.}
\bibitem{finite}{Z. Bern,{\hskip -0.1cm} J. Carrasco,{\hskip -0.1cm} L. Dixon,{\hskip -0.1cm} H. Johansson,{\hskip -0.1cm} D. Kosower{\hskip -0.1cm} and{\hskip -0.1cm} R. Roiban,{\hskip -0.1cm} {\tt hep-th}/0702112.}
\bibitem{living}{Z. Bern and A. K. Grant, {\it Phys. Lett.} {\bf B 457} (1999) 23, {\tt hep-th}/9904026. \\
Z. Bern, {\it Living Rev. Rel.} {\bf 5} (2002) 5, {\tt gr-qc}/0206071.}
\bibitem{PT}{S. J. Parke and T.R. Taylor, {\it Phys. Rev. Lett.} {\bf 56} (1986) 2459. \\
F. A. Berends and W. T. Giele, {\it Nucl. Phys.} {\bf B 306} (1988) 759.}
\bibitem{CSW}{F. Cachazo, P. Svrcek and E. Witten, {\it JHEP} {\bf 0409} (2004) 006, {\tt hep-th}/0403047.}
\bibitem{Rosly}{A. Gorsky and A. Rosly, {\it JHEP} {\bf 0601} (2006) 101, {\tt hep-th}/0510111.}
\bibitem{Mansfield}{P. Mansfield, {\it JHEP} {\bf 0603} (2006) 037, {\tt hep-th}/0511264.}
\bibitem{EM}{J. Ettle and T. Morris, {\it JHEP} {\bf 0608} (2006) 003, {\tt hep-th}/0605121. \\
J. Ettle, C. Fu, J. Fudger, P. Mansfield and T. Morris, {\tt hep-th}/0703286.}
\bibitem{QMC}{A. Brandhuber, B. Spence, G. Travaglini and K. Zoubos, arXiv:0704.0245 [\tt hep-th].}
\bibitem{ABHS}{S. Ananth, L. Brink, R. Heise and H. G. Svendsen, {\it Nucl. Phys.} {\bf B 753} (2006) 195, {\tt hep-th}/0607019.}
\bibitem{lcgref}{J. Scherk and J. H. Schwarz, {\it Gen. Rel. Grav.} {\bf 6} (1975) 537. \\
I. Bengtsson, M. Cederwall and O. Lindgren, GOTEBORG-83-55, 1983.}
\bibitem{redef}{A. Salam and J.A. Strathdee, {\it Phys. Rev.} {\bf D 2} (1970) 2869. \\
G.'t Hooft and M. Veltman, ``Diagrammar", CERN Report No.73-9, 1973. \\
H. Georgi, {\it Nucl. Phys.} {\bf B 361} (1991) 339$\;$; C. Arzt, {\it Phys. Lett.} {\bf B 342} (1995) 189.}
\bibitem{BGK}{F. A. Berends, W. T. Giele and H. Kuijf, {\it Phys. Lett.} {\bf B 211} (1988) 91.}
\bibitem{bohr}{N. E. J. Bjerrum-Bohr, D. C. Dunbar, H. Ita, W. B. Perkins and K. Risager, {\it JHEP} {\bf 0601} (2006) 009, {\tt hep-th}/0509016.}
\bibitem{sff}{L. Brink, O. Lindgren and B. E. W. Nilsson, {\it Nucl. Phys.} {\bf B 212} (1983) 401. \\
S. Ananth, L. Brink and P. Ramond, {\it JHEP} {\bf 0407} (2004) 082, {\tt hep-th}/0405150. \\
S. Ananth, L. Brink and P. Ramond, {\it JHEP} {\bf 0505} (2005) 003, {\tt hep-th}/0501079.}
\bibitem{SM}{S. Mandelstam, {\it Nucl. Phys.} {\bf B 213} (1983) 149.}
\end{thebibliography}
\end{document}